\documentclass[twocolumn, aps, pra]{revtex4-2}
\usepackage{graphicx}  
\usepackage{amsmath}   
\usepackage{amssymb}   
\usepackage{hyperref}  
\usepackage{float}
\usepackage{caption}

\begin{document}
	
	\title{Advancements in the GravAD Pipeline: Template Reduction and Testing Simulated Signals for Black Hole Detection}
	
	\author{William E. Doyle}
	\affiliation{The University of Portsmouth}
	\email{will.doyle30@gmail.com}
	\email{up928187@myport.ac.uk}
	
	\date{\today}
	
	\begin{abstract}
This paper introduces significant improvements to the GravAD pipeline, a Python-based system for gravitational wave detection. These advancements include a reduction in waveform templates, implementation of simulated signals, and optimisation techniques. By integrating these advancements, GravAD exhibits increased performance, efficiency, and accuracy in processing gravitational wave data. This leads to more efficient detection and freeing computational resources for further research. This pipeline also applies adaptive termination procedures for resource optimisation, enhancing gravitational wave detection speed and precision. The paper emphasises the importance of robust, efficient tools in gravitational wave data analysis, particularly given the finite nature of computational resources. Acknowledging system limitations such as dependency on the ripple python library capabilities and suggests future enhancements in waveform generation and differentiation. 
	\end{abstract}

	\maketitle
	
	\section{Introduction}
	
	Compact binary coalescences (CBCs), astronomical occurrences marked by the merging of two distinct compact objects such as black holes (BHs) or neutron stars (NSs), present unique opportunities to study gravitational waves (GWs) \cite{gw_sources}. Since the advent of the Laser Interferometer Gravitational-wave Observatory (LIGO) and the subsequent detection of the first GW signal in 2015 \cite{abbott_2016}, our understanding of these cosmic phenomena has significantly expanded. The resultant waveforms from CBCs, or transient-modelled waveforms, encapsulate the dynamics of these merging systems and their study enables us to probe the nature of gravity itself. Confirming the predictions of General Relativity in the strong-field regime, such as the inspiral, merger, and ringdown phases of compact object mergers, allows us to test the limits of our current understanding and potentially uncover new physics \cite{Abram_1992}.
	
	In our prior research, we introduced GravAD, a Python-based search pipeline for GW detection utilising automatic differentiation (AD) and JAX \cite{jax2018github}. GravAD's approach centres around dynamically generating and refining waveform templates, thereby improving their fit to incoming data with each iteration. This method not only enhances the efficiency of the detection process but also significantly reduces the number of templates required for data analysis \cite{Doyle_2023}.
	
	This paper aims to further expand on the advancements made to GravAD since our initial publication. We have implemented significant enhancements to the system, driven by two primary motivations. Firstly, the escalating complexity of waveform templates and the increasing sensitivity of LIGO detectors necessitate continuous improvements in data analysis methods for GWs \cite{tedwards, abbott_2020}. The growth in detector sensitivity and the rise in GW detections underscore the need for resilient and efficient analytical tools. Secondly, with the emergence of new detectors capable of observing a larger variety of CBC events, our analysis approach must become more comprehensive \cite{menge_2020}.
	
	To address these motivations, we have integrated simulated signals into our pipeline, pushing GravAD's boundaries and expanding its range of detectable astrophysical sources. We have also further decreased the number of templates needed for data analysis, thus improving the efficiency of the detection process and limiting the lost accuracy from this technique. Acknowledging the trade-off between precision and computational resource requirements \cite{Coogan_2022}, we have explored alternative optimisation algorithms. For instance, the Adam optimisation algorithm, a method that adjusts the learning rate based on the estimated moment of gradients \cite{adam}, has been partially incorporated into GravAD, yielding substantial enhancements in performance and efficiency.
	
	In this research, our primary objective is to develop our search pipeline by integrating simulated signals and innovative optimisation strategies. This includes the adoption of a callback mechanism - with the objective of early termination - reminiscent of TensorFlow's callback method \cite{tf}, which effectively halts optimisation processes once specific criteria have been fulfilled. Subsequently, we will discuss the outcomes and critically evaluate the implications of these modifications.

	\section{Methods} 
	
	The development and improvement of methodologies to detect GWs is a continually evolving field. Our research focuses on enhancing the accuracy and efficiency of the GravAD pipeline, a process that includes three primary stages: generating simulated signals, refining the optimisation strategy, and reducing the number of templates.
	
	Each of these stages has the objective of yielding high Signal-to-Noise Ratios (SNRs). The SNR is how strong the GW signal is against a background of noise. This indicator alerts us to the presence of a detection \cite{snr}. Each iteration in GravAD's search aims to refine templates based on gradient information. This works by updating mass parameters fed into the waveform generator. For more information on how GravAD works, visit our previous publication \cite{Doyle_2023}.
		
	\subsection{Optimisation Strategy Selection}
	
	In our prior endeavours, we predominantly employed stochastic gradient descent (SGD) and simulated annealing (SA) in our pursuit of detecting GWs. In a bid to further refine our search capabilities, we incorporated the concept of momentum in our gradient computations, resulting in an enhancement in the form of Adaptive Moment Estimation (Adam).
	
	The effectiveness of Adam can be attributed to its ability to combine the benefits of two extensions of SGD, specifically Root Mean Square Propagation (RMSProp) and Momentum. RMSProp employs a moving average of squared gradients to normalise the gradient, facilitating faster convergence and eliminating the risk of vanishing learning rates \cite{rmsprop}. On the other hand, Momentum takes into account past gradients to smooth out the update. Therefore, Adam, effectively mitigates the challenges of high variance in parameter updates, providing smoother convergence to optimal solutions.
	
	Despite our expectations, the implemented Adam method did not prove as effective as alternative approaches. Our findings revealed that utilising solely the momentum aspect of Adam led us to locate the optimal template more swiftly than with the comprehensive Adam application. This adjustment, importantly, maintained a higher level of accuracy than we had previously seen with GravAD.
	
	\subsubsection{Stochastic Gradient Descent}
	
	Our gradient is calculated using AD, which simplifies things greatly. We take the derivative of the SNR calculation and combine this with a learning rate in order to climb to the top of the peak, essentially performing gradient ascent.
	
	We can therefore create an updated mass parameter $\theta_i$:
	
	\begin{equation}
		\theta_i = \alpha \cdot g_i
	\end{equation}
	
	where $i$ corresponds to the index of the parameter (in this case $i=0$, corresponding to the first gradient value), and with $\alpha$ as the learning rate and $g_i$ as the current gradient.
	
	\subsubsection{Simulated Annealing}
	
	Simulated annealing complements SGD by facilitating its escape from local maxima, enhancing the optimisation process. By enabling hill-climbing moves, which may temporarily worsen the objective function value, SA offers a mechanism to explore alternative solutions in pursuit of a global optimum \cite{SA}. In GravAD we use this mechanism in the form of a perturbation.
	
	The perturbations are generated with a normal distribution and then scaled by the temperature. If $N(0,1)$ denotes a standard normal distribution (mean 0, variance 1), then the perturbations can be represented as:
	
	\begin{equation}
		P_i = T_i \cdot N(0,1)
	\end{equation}
	
	with $T_i$ as the temperature.
	
	The updated parameters are then calculated by adding the product of the learning rate and gradient (from SGD) to the perturbation (from SA). Given $\theta_i$ as the current mass (update parameter), $\alpha$ as the learning rate, and $g_i$ as the current gradient, the new mass parameter can be calculated as:
	\begin{equation}
		\theta_{i} = \theta_{i-1} + \left(\alpha \cdot g_i \right)+ P_i
	\end{equation}
	
	We commence with an initial temperature value of 1, which subsequently diminishes with each iteration due to the annealing rate of 0.99. The temperature parameter is updated by a straightforward multiplication of the current temperature with the annealing rate:
	
	\begin{equation}
		T_{i+1} = T_i \cdot \gamma
	\end{equation}
	
	Where $T_{i+1}$is the new temperature and $\gamma$ is the annealing rate.
	
	\subsubsection{Adam}
	Adam's optimisation process can be expressed through the following equations, as outlined in\cite{adam}:
	
	\begin{equation}
		m_i = \beta_1 \cdot m_{i-1} + (1 - \beta_1) \cdot g_i \label{eq:mo}
	\end{equation}
	
	where $m_i$ is the moving averages of the gradient, $g_i$ is the current gradient, and $\beta_1$ is the decay rate set to 0.9. We then calculate the gradient squared:
	
	\begin{equation}
		v_i = \beta_2 \cdot v_{i-1} + (1 - \beta_2) \cdot g_i^2
	\end{equation}
	
	where $\beta_2$ is the decay rate set to 0.999. From here we can compute the bias-corrected estimates. Firstly:
	
	\begin{equation}
		\hat{m_i} = \frac{m_i}{1 - \beta_1^i}
	\end{equation}
	
	then,
	
	\begin{equation}
		\hat{v_i} = \frac{v_i}{1 - \beta_2^i}
	\end{equation}
	
	resulting in our updated parameter:
	
	\begin{equation}
		\theta_i = \theta_{i-1} - \alpha \cdot \frac{\hat{m_i}}{\sqrt{\hat{v_i}} + \epsilon}
	\end{equation}
	
	with $\alpha$ as the learning rate, and $\epsilon$ is a small constant to avoid division by zero.
	
	\subsection{Applying the Optimisations}
	
	When we combine the optimisations of SGD, SA, and Adam we get:
	
	\begin{equation}
		\theta_i = \theta_{i-1} - \alpha \cdot \frac{\hat{m_i}}{\sqrt{\hat{v_i}} + \epsilon}  + P_i .
	\end{equation}
	
	In practice, however, this method falls short in performance when compared to the likes of SGD and SA. Prompted by this discovery, we ventured into an alternative approach, using only a segment of the Adam optimiser: the momentum.
	
	We apply equation \ref{eq:mo} in a straightforward manner. Subsequently, the updated parameter is determined by the sum of the product of the learning rate and the updated momentum (from the preceding step), and the perturbation (derived from SA). Given $\theta_i$ as the current parameter (in this context, mass1), $\alpha$ as the learning rate, and $P_i$ as the current perturbation, the new parameter is computed as follows:
	
	\begin{equation}
		\theta_i = \theta_{i-1} + \left(\alpha \cdot m_i\right) + P_i
	\end{equation}
	
	The aforementioned equations illustrate the parameter update rules when utilising SGD with momentum, combined with SA. These rules demonstrate how the gradient information, momentum, and perturbations guide the search for optimal solutions within the parameter space.
	
	The subsequent inclusion of momentum serves to further fine-tune this process. By factoring in the momentum of the gradient, GravAD facilitates swifter convergence and a more nuanced exploration of the parameter space. Consequently, the combination of these optimisation strategies not only accelerates the attainment of solutions but also boosts the likelihood of these solutions being near-optimal. This, in turn, amplifies our capabilities in GW detection.
	
	\subsection{Template Reduction Technique}
	
	GravAD implements an adaptive termination procedure to fine-tune its exploration of the parameter space, which is analogous to the callback function found in machine learning frameworks such as TensorFlow. This function is engineered to pinpoint suitable peak values in the SNR landscape, and prematurely terminate the search upon their discovery.
	
	The mechanism functions by identifying occasions when the SNR surpasses a previously recorded peak. Upon identification, the algorithm records the new peak SNR and its corresponding iteration index. If, for instance, 5 iterations pass without topping the previously recorded peak SNR, the system makes a strategic manoeuvre to conclude the search prematurely.
	
	The capability to adaptively terminate manifests as an efficient methodology for navigating the template space. This results in a substantial decrease in the number of templates required for the search. Consequently, it improves GravAD's proficiency to detect GWs rapidly.
	
\begin{figure}
		\centering
		\includegraphics[width=0.5\textwidth]{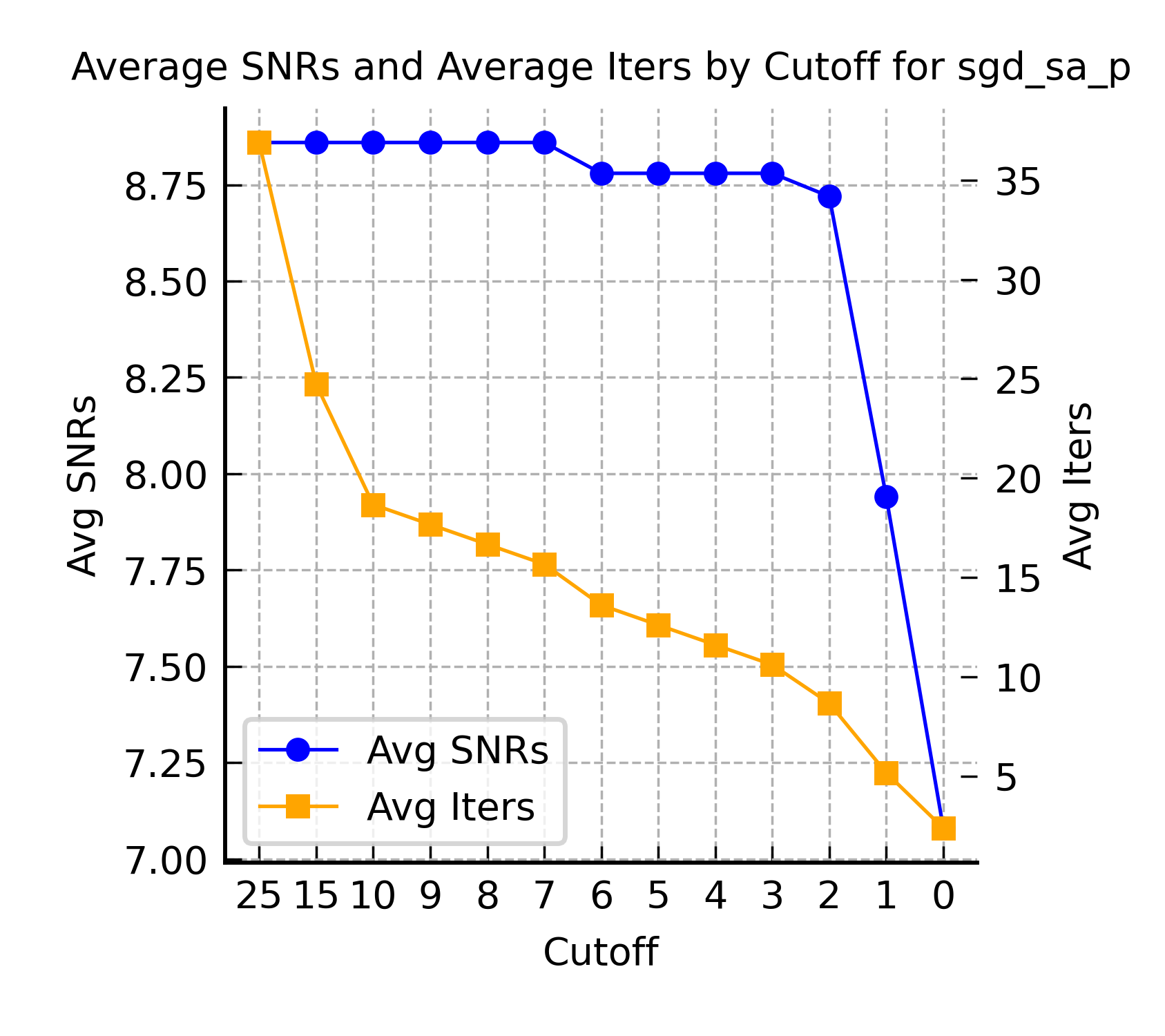}
		\caption{Trends of average SNRs and average iterations for the 'sgd\_sa\_p' optimiser, as influenced by different cutoff values.}
		\label{fig:cutoff}
	\end{figure}

	For subsequent experiments, we selected a cutoff value of 2. This indicates that if the SNR does not improve within two additional iterations, the algorithm will cease the search. This value was chosen because it doesn't significantly degrade the SNR and produces results comparable to a cutoff value of 25. It also leads to a low average number of iterations. We arrived at these values and averages by using events: GW150914, GW151012, GW151226, GW170104, GW170729, GW170809, GW170814, GW170818, GW170823. The outcomes of these tests are visualised in Figure \ref{fig:cutoff}.
	
		\subsection{Simulated Signals Generation}
	
	Simulated signals, defined as synthetic data crafted to mimic real GWs, prove instrumental in testing the effectiveness of the GravAD pipeline. They provide valuable insights into the accuracy of the algorithm by facilitating comparisons between known parameters and those estimated.
	
	The python library, \texttt{ripple} \cite{tedwards}, is utilised to generate these simulated signals according to predetermined parameters, amid the interference of noise. This inclusion of noise is a methodological decision intended to assess the efficacy of GravAD when applied to realistic signals. Each simulated signal is created with a pair of masses. Masses from 20 to 100 (with a step of 10) are used for both the primary and secondary bodies involved in the simulated gravitational event. This consequently results in the generation of signals that correspond to the coalescence of two objects, one of which, for example, could possess a mass of 20 solar masses, while the other could potentially be as massive as 80 solar masses. For each pair of masses, the frequency domain waveform of the GW signal is generated using the \texttt{gen\_waveform} function from the GravAD library. This function takes as input the masses of the two bodies, a frequency series determined by a delta frequency/step size, and a set of parameters describing the spins, distance, and phase; however, these are all set to the same value as they have minimal impact on the search.
	
	Once the waveform is generated, it is then transformed into a noisy signal in the frequency domain. This process is proceeded by adding a noise profile obtained from the Power Spectral Density (PSD) derived from the event 'GW150914' detected by the LIGO Hanford detector (H1).
	
	The noisy waveform, $h_{noisy}(f)$ is given by: 
	
	\begin{equation}
		h_{noisy}(f) = h(f) + N(f) 
	\end{equation}
	
	Where $h(f)$ is the generated waveform, $N(f)$ is the noise profile from the PSD.
	
	This approach to generating and storing a wide range of simulated signals with varying mass parameters allows for robust testing of the GravAD pipeline under different signal scenarios.
	
	\section{Results} 
		
	\subsection{Effectiveness of the Optimisations}
	
\begin{figure}
		\centering
		\includegraphics[width=0.5\textwidth]{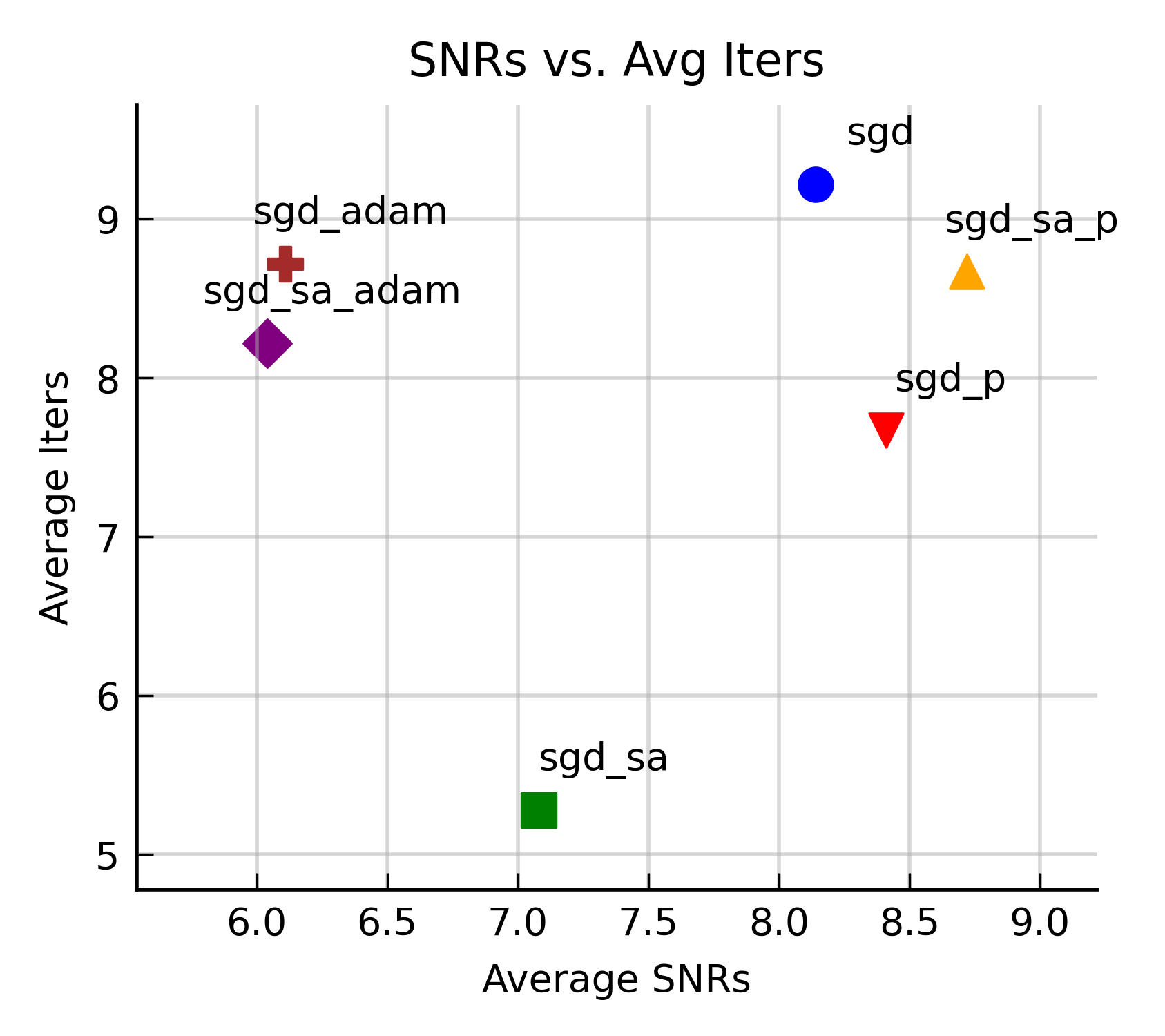}
		\caption{A comparison between different optimisers used on GravAD.}
		\label{fig:optims}
	\end{figure}

	Upon comparing the performance of various optimisation techniques (as depicted in Figure \ref{fig:optims}), we observe that standalone SGD can be somewhat slow yet effective, resulting in a high SNR. The incorporation of SA navigates the search away from local maxima, guiding it towards regions within the parameter space that are more likely to generate superior solutions. While this approach diminishes the average iterations per run, it also adversely affects the average SNR. The most effective strategy, on the whole, appears to be the combination of SGD, SA, and momentum (P), due to its high average SNR and reasonable average iterations. We can also see the ineffectiveness of Adam in our search with and without SA involvement. 
	
	\subsection{Significant Reduction in Template Usage}
	
	Our refined methodology yielded a considerable reduction in the number of templates necessary for performing a search. Traditionally, an estimated $N \sim 500,000$ templates are utilised for such a task \cite{temp_num}. Our prior research, however, was able to cut down this number to a mere $N \sim 180$ templates. The current implementation advances this further, demanding on average only $N \sim 8.67$ templates per search, which signifies a remarkable efficiency gain, reducing the requirement by approximately $N \sim 60,000$ times.
	
	\subsection{Achieved SNRs}

\begin{figure}
	\centering
	\includegraphics[width=0.5\textwidth]{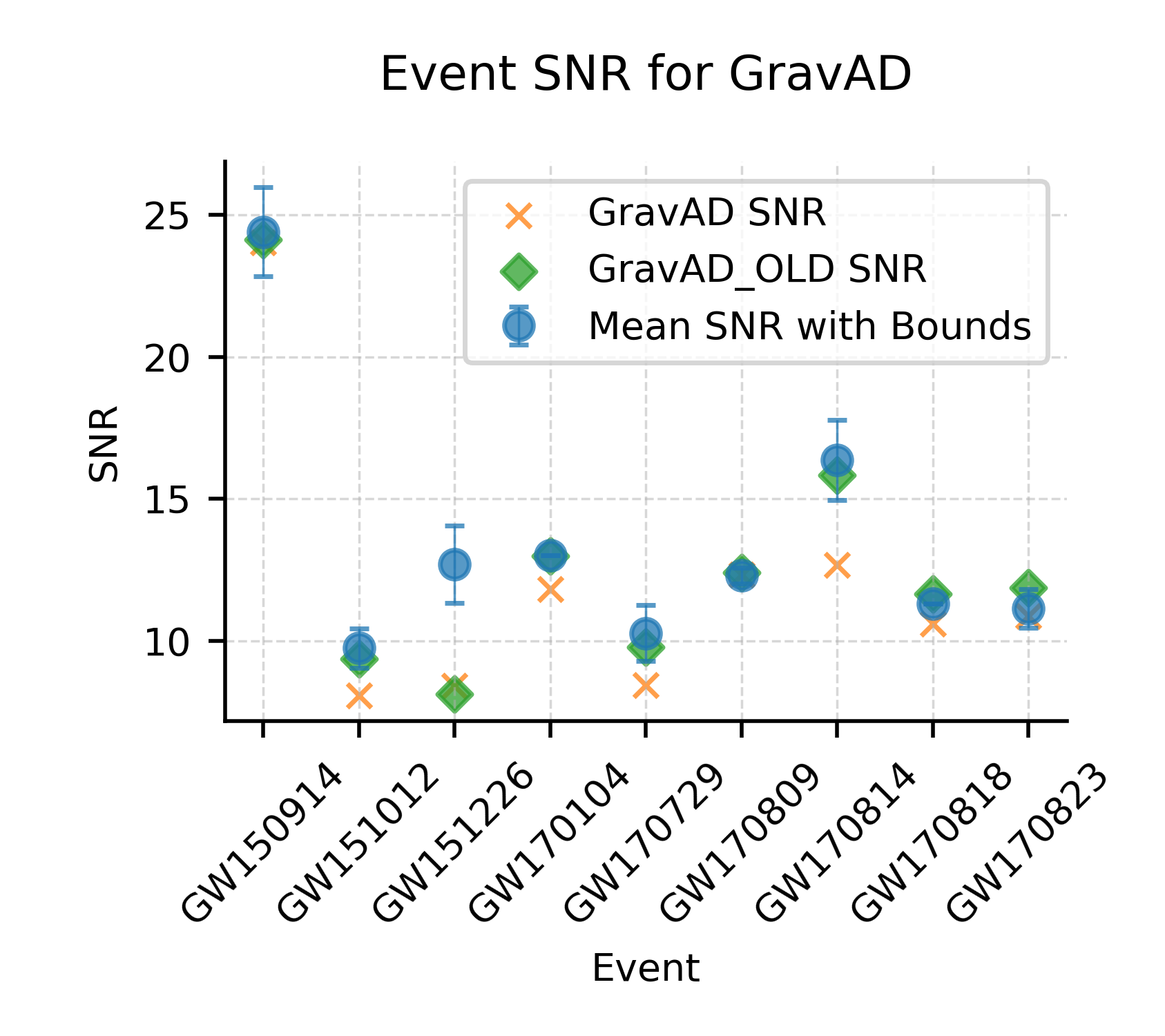}
	\caption{A comparison between the latest development of GravAD ($N \sim 8.67$ templates) compared to the predecessor ($N \sim 180$ templates), the primary differences are due to the early termination template reduction technique as well as a momentum optimiser. Also, the mean SNR with bounds data being from the GWTC \cite{gwtc}.}
	\label{fig:snr_comps}
\end{figure}

Upon reviewing the updates made to GravAD in this latest iteration, we observe (as depicted in Figure \ref{fig:snr_comps}) an average $8\%$ decrease in the magnitude of the SNR. This compromise in detection capabilities is directly linked to the minimal number of templates employed in the search.

Nonetheless, the architecture of the GravAD code provides a significant advantage. It allows researchers to harness its inherent flexibility, notably by modifying the early termination process to improve SNR values, albeit at the expense of employing more templates. This adaptability empowers researchers to strike a balance that best serves their specific research needs.

	\subsection{Performance on Simulated Signals}

Analysis of the generated signals revealed some variations in the predicted and actual mass parameters. For example, the signal simulated with a mass1 and mass2 of 90\_20 indicated a noticeable overestimation of mass1 while mass2 was underestimated. Despite this discrepancy, the total mass maintained a marginal deviation from the expected values, underscoring the balance between the mass estimations.

The precision exhibited across all simulated signals was remarkable, achieving a near-perfect alignment within $0.3\%$ of the forecasted values. This achievement validates GravAD's efficacy in accurately processing a diverse range of mass ratios. Notably, the signal with a 50\_40 mass ratio presented an anomaly, achieving only $35\%$ of the expected value.

\begin{figure*}[b]
	\centering
	\includegraphics[width=\textwidth]{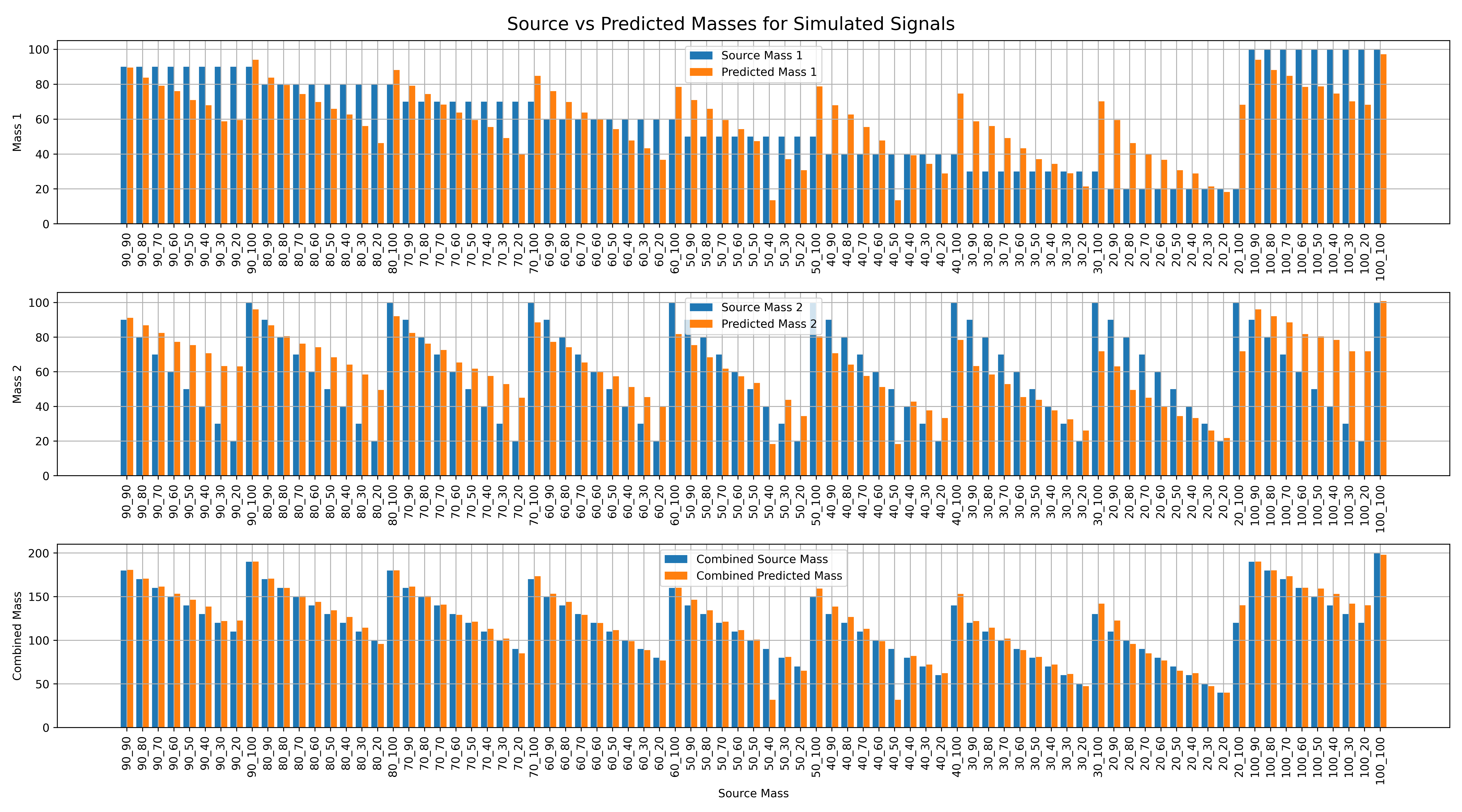}
	\caption{This graph contrasts the source and predicted masses for simulated signals, displaying distinct bars for Mass 1, Mass 2, and their combined values. An event where the orange bar (indicating predicted mass) closely aligns with the blue bar (representing source mass) signifies an accurate prediction made by GravAD.}
	\label{fig:sim}
\end{figure*}

This reinforces our confidence in GravAD's ability to process real GW signals effectively. Figure \ref{fig:sim} illustrates GravAD's competence in processing an array of mass parameters, including high and low mass ratios, and simulated data sets. This proficiency underpins GravAD's robustness and wide-ranging applicability. 
	
	\section{Discussion}
	
	The discussion section aims to provide a comprehensive analysis of the advancements made to the GravAD system, focusing on its rapid search functionality and accuracy in detecting GWs. By combining various techniques, including the integration of simulated signals and optimisation strategies, the pipeline demonstrates its effectiveness in efficiently processing GW data. 
	
	One significant aspect to consider is the importance of efficient methods in allocating computational resources for other research. As the field of GW detection continues to evolve and the number of detections increases, computational hardware becomes a valuable and limited resource. The GravAD system addresses this challenge by implementing optimisation techniques that reduce the number of templates required in the search, allowing for the allocation of computational resources in other research areas.
	
	However, it is crucial to acknowledge the limitations of GravAD. The system is limited by the capabilities of the ripple software. This limitation implies that the effectiveness of our algorithm is dependent on the capabilities and advancements of the software it relies on. Therefore, future improvements in waveform generation and differentiation will play a crucial role in enhancing the effectiveness of the GravAD pipeline.
	
	By integrating simulated signals and optimisation strategies, the algorithm is more effective at processing GW data. The optimisation techniques, such as the use of a callback method, steer the search away from the un-needed exploration of the parameter space, improving the efficiency of the search process.
	
	Moreover, the GravAD system's ability to accurately process a diverse range of mass ratios, as demonstrated in the simulated signals, reinforces its credibility.

	\section{Conclusion}
	
	This study underscores the notable advancements in GravAD's functionality and efficiency in detecting gravitational waves. Leveraging a multitude of techniques, the system has improved its capability to process simulated signals, thereby enhancing the accuracy of gravitational wave detection. Despite minor discrepancies in individual mass predictions, GravAD adeptly preserved total mass values, showcasing its practicality across diverse mass ratios.
	
	Our research also highlights the vital role of optimisation strategies in augmenting the efficacy and speed of GravAD's search process. The blend of SGD, SA, and momentum has proven effective, offering a balance between high average SNR and reasonable average iterations of $N\sim8.67$ for each search.
	
	A pivotal achievement lies in GravAD's substantial reduction in the number of templates required for search processes, improving computational efficiency remarkably without substantial reduction in result quality. This major stride forward paves the way for the allocation of resources to other vital research areas, proving instrumental in the continual expansion of gravitational wave detection.
	
	The success of GravAD remains tethered to the progression of the ripple library. Future enhancements in the generation and differentiation of waveforms could further boost GravAD's effectiveness. Therefore, the continual evolution of our system necessitates paralleled advancements in the underlying technology.
	
	\bibliographystyle{unsrtnat}  
\bibliography{bib}  

	\begin{acknowledgments}
		I would like to thank Ian Harry for his endorsement to publish this paper.
	\end{acknowledgments}

	\appendix
	\section{}
	For the full GravAD code, visit: \url{https://github.com/WDoyle123/GravAD}
	
\end{document}